\documentclass[aps,prb,reprint,superscriptaddress,showpacs]{revtex4-1}

\usepackage{graphics}

\begin{document}

\title{%
Absence of a classical long-range order
in $S=1/2$ Heisenberg antiferromagnet
on triangular lattice}

\author{Nobuo Suzuki}
\email[]{nobu@ait.tbgu.ac.jp}
\affiliation{%
Faculty of Science and Technology, %
Tohoku Bunka Gakuen University, %
Sendai 980-8551, Japan }

\author{Fumitaka Matsubara}
\affiliation{%
Department of Applied Physics, %
Tohoku University,
Sendai 980-8579, Japan }

\author{Sumiyoshi Fujiki}
\affiliation{%
Faculty of Science and Technology, %
Tohoku Bunka Gakuen University, %
Sendai 980-8551, Japan %
}%

\author{Takayuki Shirakura}
\affiliation{%
Faculty of Humanities and Social Sciences, %
Iwate University, Morioka 020-8550, Japan  %
}%

\date{\today}

\begin{abstract}

We study the quantum phase transition
  of an $S=1/2$ anisotropic $\alpha$ $(\equiv J_z/J_{xy})$ 
 Heisenberg antiferromagnet
 on a triangular lattice. 
We calculate the sublattice magnetization
 and the long-range helical order-parameter
 and their Binder ratios on 
 finite systems with $N \leq 36$ sites. 
The $N$ dependence of the Binder ratios reveals
 that the classical 120$^{\circ}$ N\'{e}el state occurs
 for $\alpha \lesssim 0.55$,
 whereas a critical collinear state occurs
 for $1/\alpha \lesssim 0.6$. 
This result is at odds with a widely-held belief
 that the ground state of a Heisenberg antiferromagnet is
 the 120$^{\circ}$ N\'{e}el state,
 but it also provides a possible mechanism
 explaining experimentally observed spin liquids. 

\end{abstract}


\maketitle


Because an exotic spin state may occur
 as a result of low-dimensional quantum fluctuations
 and geometric frustration, 
 the $S = \frac{1}{2}$ quantum antiferromagnetic Heisenberg (QAFH) model 
 on the triangular lattice is one of the central issues
 in solid-state physics.
Anderson proposed a resonating-valence-bond (RVB) state or 
 a spin-liquid (SL) state as the ground state (GS).\cite{anderson1973}
Since then, many theoretical studies have focused on identifying
 the GS by using different methods
 such as spin-wave (SW) theory,\cite{miyake1992} 
 variational Monte Carlo techniques,\cite{huse1988,sindzingre1994}
 series expansions,\cite{singh1992,elstner1993} 
 exact diagonalizations (ED) of finite systems,\cite{%
 fujiki1987,*fujiki1987a,*fujiki1986,
 nishimori1988,*nishimori1989a,*nishimori1989b,
 bernu1992,*bernu1994,leung1993,richter2004} 
 quantum Monte Carlo techniques,~\cite{capriotti1999}
 density matrix renormalization group theory,\cite{white2007}
 and diagrammatic Monte Carlo techniques.\cite{kulagin2013}
The GS is now widely believed to be a long-range-order (LRO) state
 with the 120$^{\circ}$ sublattice structure
 (the 120$^{\circ}$ N\'{e}el state)
 because the results of most numerical studies can be analyzed
 by using this image.\cite{elstner1993,bernu1994,capriotti1999,richter2004}
However, experimental developments have enabled us
 to synthesize model compounds such as 
 $\kappa$-(ET)$_2$Cu$_2$(CN)$_3$,\cite{shimizu2003}
 EtMe$_3$Sb[Pd(dmit)$_2$]$_2$,\cite{itou2008}
 and Ba$_3$IrTi$_2$O$_9$.\cite{tusharkanti2012}
In these compounds,
 no spin ordering has been observed down at very low temperatures;
 several mechanisms have been proposed to resolve this discrepancy,
 such as spatial anisotropy,\cite{morita2002,yunoki2006}
 ring exchange,\cite {motrunich2005}
 and spinon interaction.~\cite{lee2007}
 
Before examining these mechanisms,
we must first carefully re-examine the GS properties of the QAFH model
 because the base of the 120$^{\circ}$ N\'{e}el GS
 is not yet solidly established. 
In particular,
 even in the most widely accepted studies,
 the magnitude of the sublattice magnetization (SMAG) $m^{\dagger}$
 is not compatible. 
SW theory in finite systems \cite{bernu1994}
 and the quantum Monte Carlo technique \cite{capriotti1999}
 suggest $m^{\dagger} = 0.4\sim0.5$
 in the classical case
 units of $m^{\dagger} = 1$,
 whereas numerical series expansions suggest either 
$m^{\dagger} \sim 0$ \cite{singh1992} or some small value.\cite{elstner1993}
In the ED technique up to $N = 36$ spins, 
results depend on the scaling functions,
 which gives either $m^{\dagger} \sim 0.5$ 
\cite{bernu1992,richter2004} or $m^{\dagger} \sim 0$.\cite{leung1993}
The quantum Monte Carlo technique \cite{capriotti1999}
 does not  satisfactorily reproduce ED results for $N = 12$ and 36.

In the present paper, we report that the GS of the QAFH model
 differs from the 120$^\circ$ N\'{e}el state. 
We consider finite systems with $N$ $(\leq 36)$ sites in the usual way,
 but take a different approach. 
To investigate the quantum phase transition,
 we consider with an anisotropic model.
We calculate the SMAG and the long-range helical order (LRHO) parameter
 and examine the Binder ratios of these quantities. 
We find that, in concurrence with recent results,
 the GS is a critical state
 with collinear structure in the Ising-like range
 and a 120$^{\circ}$ N\'{e}el state in the XY-like range. 
In contrast, the GS is a SL state
 in the Heisenberg-like range. 
We estimate an anisotropy threshold
 for the occurrence of the critical state
 and for the 120$^{\circ}$ N\'{e}el state. 


We start with an anisotropic model on periodic finite lattices
 described by the Hamiltonian 
\begin{eqnarray}
{\cal H} = 2J\sum_{\langle i,j \rangle }%
 [S^x_iS^x_j+S^y_iS^y_j + \alpha S^z_iS^z_j],
\label{hamiltonian}
\end{eqnarray}
where $J > 0$, $\alpha \ge 0$, 
 and the sum runs over all the nearest-neighbor pairs of sites. 
Note that the model with $\alpha = \infty$ is an Ising model 
for which the GS is a critical state
characterized by a power low decay
 of the spin correlation function.~\cite{stephenson1964} 
At the other limit,
 the model with $\alpha \sim 0$ is an XY-like 
model for which the 120$^{\circ}$ N\'{e}el state is suggested to 
occur.\cite{leung1993,suzuki1995} 
We discuss the spin structure of the Heisenberg-like model 
with $\alpha \sim 1$ by comparing the properties of this model 
with those of the Ising- and XY-like models. 
The main issue is whether $m^{\dagger} \neq 0$ or not. 

\begin{figure}
\includegraphics{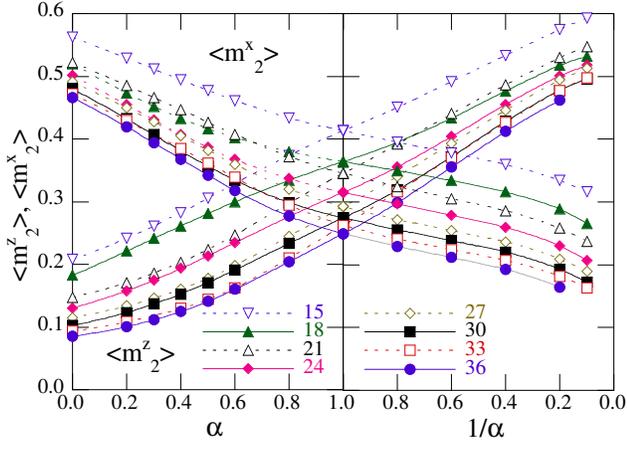}%
\caption{\label{fig:allmag} (Color online)
$z$ and $xy$ components of the SMAG 
 ($\langle m_2^z\rangle_N$ and $\langle m_2^{x}\rangle_N (\equiv 
 \langle m_2^{xy}\rangle_N$/2), 
 respectively)
in the GS as functions of $\alpha$. }
\end{figure}

By using a power method,
 we calculate the GS eigenfunction $|\psi_G \rangle$
 for two types of lattices, A and B,
 with $N \leq 36$ sites.
Type-A lattices have $N = 9$, 12, 21, 27, 36, 
and type-B lattices have $N = 15$, 18, 24, 30, 33. 
The shapes of the type-A lattices were
 presented in Ref.~\onlinecite{leung1993};
 for this lattice type,
 the sublattices $\Omega_1$,  $\Omega_2$, and $\Omega_3$ are equivalent. 
The type-B lattices are constructed so that
 the 120$^{\circ}$ N\'{e}el structure is possible in the classical case. 
The SMAG of the type-A lattices,
 and in particular their $N$ dependence,
 have already been studied
 by several groups.~\cite{bernu1992,bernu1994,leung1993, richter2004}
However, for these small systems,
 the data strongly depend on the parity and magnitude of $N$.
In the present work,
 we add to these the data for type-B lattices.


First, we consider the SMAG. The $\nu$ component 
($\nu = x, y, z$) of the square of the magnetization 
of the $\Omega_l$ sublattice is defined as
\begin{eqnarray}
m_{l}^{\nu} &=&  \frac{1}{(N/6)^2}(\sum_{i \in \Omega_l}S_i^{\nu})^2,
\end{eqnarray}
and the $xy$ component is defined as $m_{l}^{xy}=m_{l}^{x} + m_{l}^{y}$. 

Figure \ref{fig:allmag} shows the $xy$ and $z$ components
 of the SMAGs $\langle m_2^{\mu} \rangle_N 
(\equiv \frac{1}{3}\sum_l^3\langle m_{l}^{\mu} \rangle_N )$ 
$(\mu = z, xy)$ as functions of $\alpha$,
 where $\langle A \rangle_N$ = $\langle \psi_G| A(N) |\psi_G \rangle$. 
For $\alpha \sim 0$,  $\langle m_2^{xy} \rangle_N$ has a large value
 and is only weakly dependent on size,
whereas $\langle m_2^z \rangle_N$ is small
 and depends strongly on size.
As $\alpha$ increases,
 $\langle m_2^{xy} \rangle_N$ gradually 
 decreases and $\langle m_2^z \rangle_N$ increases,
 and $\langle m_2^z \rangle_N = \langle m_2^{xy} \rangle_N/2$ 
at $\alpha = 1$. 
The reverse is true for $1/\alpha \sim 0$. 
The results at $\alpha \sim 0$ and $1/\alpha \sim 0$ seem to be 
compatible with the classical picture of the GS. 
However, in contrast with the classical case, 
$\langle m_2^z \rangle_N$ (or $\langle m_2^{xy} \rangle_N$)
 does not abruptly increase (or decrease) as $\alpha$
 is increased across the Heisenberg point $\alpha = 1$. 


We now examine the quantum phase transition of the model 
 by considering the dependence of $\alpha$ on
$\langle m_2^z\rangle_N$ and $\langle m_2^{xy}\rangle_N$.  
The SMAG at $\alpha = 1$ for 
$N \rightarrow \infty$ has been estimated
 by several groups\cite{bernu1992,leung1993,richter2004}
 who used different scaling relations. 
However, the result depends on both the units of the sublattice 
magnetization and the scaling functions. 
Here we consider the Binder ratios\cite{binder1982} of 
$\langle m_2^z\rangle_N$ and $\langle m_2^{xy}\rangle_N$ which are 
free from the scaling function and their units.
The Binder ratios of $\langle m_2^z\rangle_N$ and 
$\langle m_2^{xy}\rangle_N$, $B_m^{z}(N)$ and $B_m^{xy}(N)$, 
respectively, are defined as 
\begin{eqnarray}
B_m^{z}(N) &=&(3- \langle m^{z}_4\rangle_N/\langle m_2^{z}\rangle_N^2)/2,\\
B_m^{xy}(N)&=&(5- 3\langle m^{xy}_4\rangle_N/\langle m_2^{xy}\rangle_N^2)/2,
\end{eqnarray}
where  $\langle m_4^{\mu}\rangle_N \equiv \frac{1}{3}\sum_l^3 
\langle \psi_G|(m_{l}^{\mu})^2 |\psi_G \rangle$.

\begin{figure}
\includegraphics{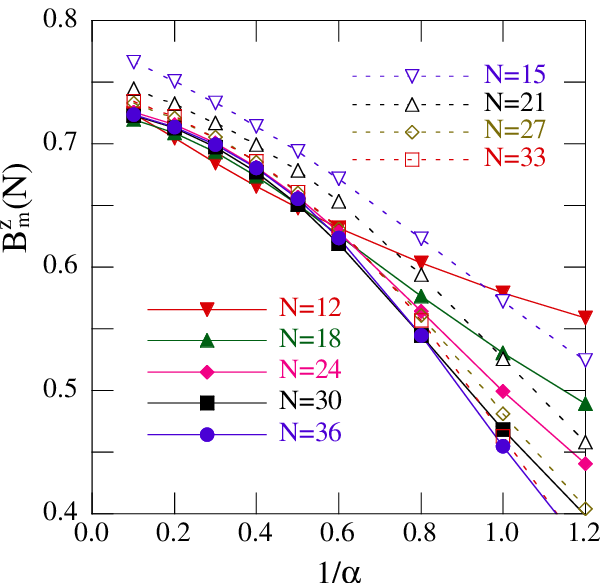}%
\caption{\label{fig:zbind_a} (Color online)
Binder ratios  $B_m^{z}(N)$ as functions of $1/\alpha$. 
The ratios for $N$ even and odd are shown by solid and open symbols, 
respectively.
}
\end{figure}
\begin{figure}
\includegraphics{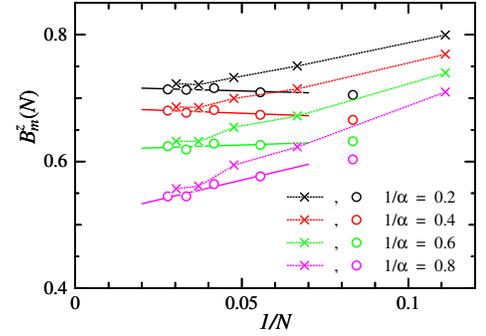}%
\caption{\label{fig:zbind_N} (Color online)
Binder ratios $B_m^{z}(N)$ for different $1/\alpha$
 as functions of $1/N$.
 Ratios for $N$ even and odd are shown 
by circles and crosses, respectively.
The straight lines for $N$ even are the least square fits for $N \geq 18$.
}
\end{figure}

We first examine the GS of the Ising-like model for $1/\alpha < 1$. 
In Fig.~\ref{fig:zbind_a}, we plot  $B_m^{z}(N)$ as functions of 
$1/\alpha$.  
The dependence of $B_m^{z}(N)$ on $N$
 differs somewhat for $N$ odd or even. 
For $N$ even, $B_m^{z}(N)$ at $1/\alpha \sim 1$ decreases
 with increasing $N$,
 revealing that $\langle m_2^z\rangle_N$ vanishes
 as $N \rightarrow \infty$. 
As $1/\alpha$ decreases, $B_m^{z}(N)$ for different $N$ increase, 
come together at $1/\alpha \sim 0.6$,
 and then gradually increase thereafter. 
This result is consistent with the fact that the GS is critical 
at $1/\alpha = 0$.\cite{stephenson1964} 
For $N$ odd, although $B_m^{z}(N)$ are larger than for $N$ even,
 even at $1/\alpha \sim 0$ they decreases with increasing $N$. 
To resolve this discrepancy,
 we show in Fig.~\ref{fig:zbind_N} a plot of $B_m^{z}(N)$
 as functions of $1/N$.
We see that, as $N$ increases,
 $B_m^{z}(N)$ for odd $N$ approaches to those for even $N$. 
Thus, we conclude that the decrease of $B_m^{z}(N)$ for small $N$
 is an abnormal finite-size effect that comes from the difference 
in the ratio $r_z = M_z/N$,
 with $M_z$ being the $z$ component 
of the total-spin number.\cite{abnormal}
The slopes of the fitting lines of $B_m^{z}(N) vs. 1/N$
shown in Fig.~\ref{fig:zbind_N} are almost zero
 for $1/\alpha \lesssim 0.6$.
We suggest that the GS is the critical state for 
$\alpha > \alpha_c^z$ with $1/\alpha_c^z \sim 0.6$.

\begin{figure}
\includegraphics{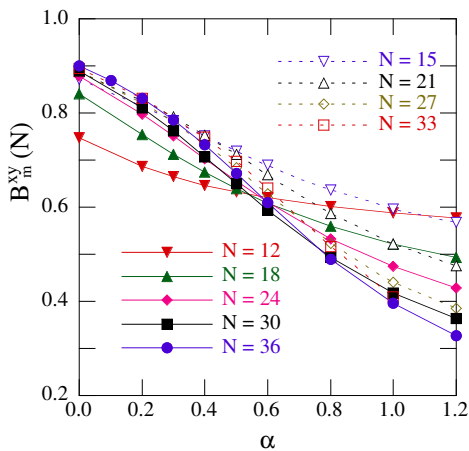}%
\caption{\label{fig:xybind_a} (Color online)
Binder ratios  $B_m^{xy}(N)$ as functions of $\alpha$. 
Ratios for $N$ even and odd are shown by solid and open symbols, 
respectively.
}
\end{figure}

\begin{figure}
\includegraphics{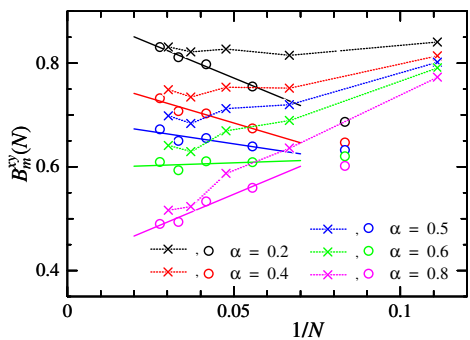}%
\caption{\label{fig:xybind_N} (Color online)
Binder ratios  $B_m^{xy}(N)$ for different $\alpha$ 
as functions of $1/N$. 
Ratios for $N$ even and odd are shown
by circles and crosses, respectively.
The straight lines for $N$ even are the least square fits for $N \geq 18$.
}
\end{figure}

Next we examine the GS of the XY-like model for $\alpha < 1$. 
Figures \ref{fig:xybind_a} and \ref{fig:xybind_N}
 show plots of $B_m^{xy}(N)$ as functions of $\alpha$ 
and of $1/N$, respectively. 
We see in Fig.~\ref{fig:xybind_N} that $B_m^{xy}(N)$ for $N$ odd also exhibit 
the abnormal finite-size effect;
 they take on values larger than those for $N$ even,
 and approach the $N$-even values as $N$ increases.
We thus consider the dependence of $B_m^{xy}(N)$ on $N$ for $N$ even. 
At $\alpha \sim 0$, $B_m^{xy}(N)$ increases with $N$. 
This result is consistent with the recently reported presence of the LRO
 in the XY model.\cite{leung1993}
However, at $\alpha \sim 1$, $B_m^{xy}(N)$ decreases with increasing $N$,
 which reveals that $\langle m_2^{xy}\rangle_N$ vanishes 
 as $N \rightarrow \infty$. 
The most remarkable point is that $B_m^{xy}(N)$ for different $N$ 
cross at $\alpha \sim 0.55$ (see also Fig.~\ref{fig:xybind_N}). 
Thus, we suggest that a quantum phase transition between 
the SL state and the LRO state 
occurs at $\alpha = \alpha_c^{xy} ( \sim 0.55)$. 


\begin{figure}
\includegraphics{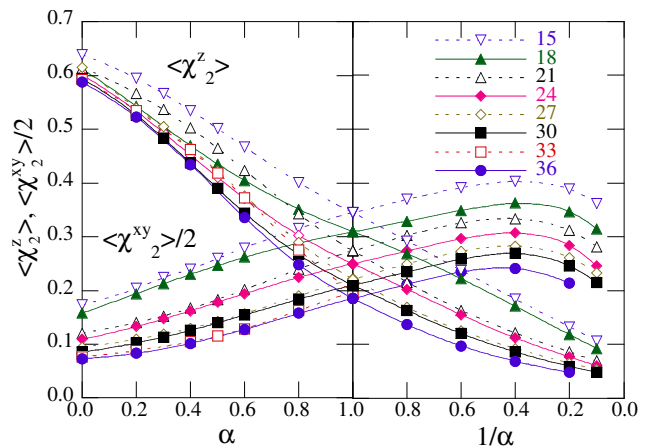}%
\caption{\label{fig:allchi} (Color online)
$z$ and $xy$ components of LRHO parameter
 ($\langle\chi_2^z\rangle_N$ and 
$\langle\chi_2^{xy}\rangle_N$, respectively)
 in the GS as functions of $\alpha$. }
\end{figure}

We now consider the helicity,
 which gives a complementary view of the spin ordering
 (i.e., it is sensitive to the $120^{\circ}$ structure). 
The local helicity \cite{fujiki1987} for each upright triangle
 at $\vec R$ is defined by 
\begin{eqnarray}
\vec{\chi}(\vec{R}) = \frac{2}{\sqrt{3}}(\vec{S}_i \times \vec{S}_j 
        + \vec{S}_j \times \vec{S}_k + \vec{S}_k \times \vec{S}_i). 
\end{eqnarray}
The order of $i \rightarrow j \rightarrow k$ is counterclockwise.
The LRHO parameter in the $\nu$ component is defined as
\begin{eqnarray}
\chi_2^{\nu} &=& \frac{1}{N^2}( \sum_{\vec{R} \in \Delta} 
             \chi^{\nu}(\vec{R}))^2, 
\end{eqnarray}
where the sum is over all upright triangles. 
We consider the LRHO parameter in the $xy$ plane, $\chi_2^z $, 
and in a plane orthogonal to the $xy$ plane
 (hereinafter called the $yz$ plane), 
$\chi_2^{xy} (= \chi_2^{x}+ \chi_2^{y})$. 
Note that $\chi_2^z $ was already calculated by several 
authors.\cite{fujiki1987,leung1993,suzuki1995} 
Here we add $\chi_2^{xy}$ to examine the occurrence of 
a distorted 120$^{\circ}$ structure in the $yz$ plane.
In the classical case, 
$\chi_2^z = 1$ and $\chi_2^{xy} =0$ for $0 \le \alpha < 1$, 
whereas $\chi_2^z = 0$ and $\chi_2^{xy} \lesssim 1$
 for $1/\alpha \lesssim 1$
 (i.e., $\chi_2^z$ and $\chi_2^{xy}$ 
suddenly exchange their role at $\alpha = 1$).

Figure \ref{fig:allchi} shows $\langle\chi_2^z\rangle_N$ and 
$\langle\chi_2^{xy}\rangle_N$ as functions of $\alpha$. 
We see that $\langle\chi_2^z\rangle_N$ has properties similar 
to those of $\langle m_2^{xy}\rangle_N$: it takes on a large value 
at $\alpha \sim 0$ and decreases with increasing $\alpha$. 
However, the dependence of $\langle\chi_2^{xy}\rangle_N$
 on $\alpha$ differs somewhat from that of 
$\langle m_2^{z}\rangle_N$; although it increases with $\alpha$, 
its increment is suppressed for $\alpha > 1$ ($1/\alpha < 1$). 
In particular,
 it reaches a maximum at $1/\alpha \sim 0.4$ and then decreases.
This is a consequence of the spin state becoming 
collinear at the Ising limit $1/\alpha \rightarrow 0$. 
Note that, even for $1/\alpha \sim 0.4$,
 $\langle\chi_2^{xy}\rangle_N$ depends strongly on $N$,
 which reveals the absence of the $xy$-component LRHO in this model. 
That is, the critical state for $\alpha > \alpha_c^z$ has a 
collinear spin structure along the $z$-axis. 
A remarkable point is that, like $\langle m_2^z\rangle_N$ and 
$\langle m_2^{xy}\rangle_N$, $\langle\chi_2^z\rangle_N$ and 
$\langle\chi_2^{xy}\rangle_N$ for $\alpha < 1$ are smoothly 
connected with those for $\alpha > 1$. 
This result supports the finding above that
 the spin structure does not changes abruptly
 at the Heisenberg point $\alpha = 1$.

\begin{figure}
\includegraphics{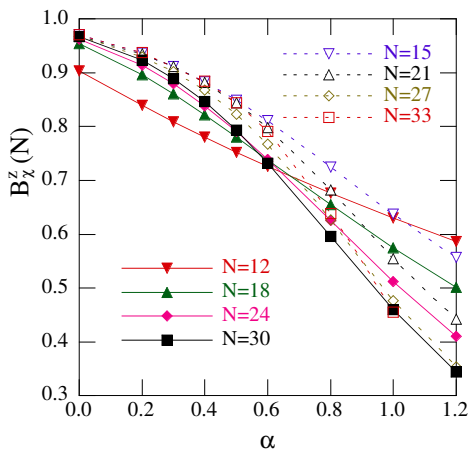}%
\caption{\label{fig:bz_chi} (Color online)
Binder ratios $B_{\chi}^z$ as functions of $\alpha$. 
Ratios for $N$ even and odd are shown by solid and open symbols,
 respectively.
}
\end{figure}

To examine the presence of the 120$^\circ$ structure in the $xy$ plane,
 we consider the Binder ratio $B_{\chi}^z(N)$ of 
$\langle\chi_2^{z}\rangle_N$, which is defined as 
\begin{eqnarray}
B_{\chi}^z(N) 
    = (3- \langle\chi^z_4\rangle_N/\langle\chi_2^z\rangle_N^2)/2.
\end{eqnarray}
Figure \ref{fig:bz_chi} shows plots of $B_{\chi}^z(N)$
 as functions of $\alpha$. 
We see that $B_{\chi}^z(N)$ exhibit properties quite similar to 
$B_{m}^{xy}(N)$;
 the abnormal finite-size effect of $B_{\chi}^z(N)$ 
 for $N$ odd,
 at $\alpha \sim 1$
 $B_{\chi}^z(N)$ is smaller as $N$ increases,
 and  at $\alpha \sim 0$ the reverse is true. 
The most interesting point is that 
$B_{\chi}^z(N)$ for different $N$ even intersect at $\alpha \sim 0.6$.
This value of $\alpha \sim 0.6$ is consistent 
with the critical value $\alpha_c^{xy} \sim 0.55$
 that is estimated from $B_{m}^{xy}(N)$. 
That is, the LRHO accompanies the LRO of the SMAG. 
Thus, we conclude that a quantum phase transition from 
the SL state to the 120$^{\circ}$ N\'{e}el state occurs at 
$\alpha = \alpha_c^{xy} \sim 0.55$. 
We should note, however, that further studies are necessary
to establish the critical value of $\alpha_c^{xy}$
as well as that of $\alpha_c^{z}$.

We thus studied the GS property of the anisotropic quantum 
antiferromagnetic Heisenberg (QAFH) model on a finite triangular 
lattice  with $N \le 36$ sites. 
We find that the GS of the model is the 120$^{\circ}$ N\'{e}el 
state for $\alpha < \alpha_c^{xy} ( \sim 0.55)$
 and is the critical collinear state
 for $1/\alpha < 1/\alpha_c^z ( \sim 0.6)$. 
That is, classical LRO is absent at $\alpha \sim 1$. 
Although this result contrasts strongly with recent theoretical ideas,
 it is consistent with recent experiments. 
We hope that our results will stimulate both 
theoretical and experimental works
 in low-dimensional frustrated quantum systems.

\begin{acknowledgments}

Part of the results in this research was obtained
 using the supercomputing resources 
at Cyberscience Center, Tohoku University. 
\end{acknowledgments}

\bibliography{suzuki}

\begin{thebibliography}{30}
\bibitem [{abn()}]{abnormal}%
  \BibitemOpen
  \href@noop {} {}\bibinfo {note} {As $r_z$ increases, $B_m^z$ increases
  because the number of basis states decreases. For $N$ odd, $r_z$ $(= 1/2N)$
  decreases with increasing $N$, which weakens the effect, and it is absent for
  $N$ even.}\BibitemShut {Stop}%
\end{thebibliography}

\end{document}